\documentclass
[twocolumn,showpacs,preprintnumbers,amsmath,amssymb,prc]{revtex4}
\usepackage{graphicx}
\usepackage{dcolumn}
\usepackage{bm}

\begin{document}
\title{A hybrid model for studying nuclear multifragmentation around Fermi energy domain: Case for central collision of Xe on Sn}

\author{S. Mallik$^1$, G. Chaudhuri$^1$ and S Das Gupta$^{2}$}

\affiliation{$^1$Theoretical Physics Division, Variable Energy Cyclotron Centre, 1/AF Bidhan Nagar, Kolkata 700064, India}
\affiliation{$^2$Physics Department, McGill University, Montr{\'e}al, Canada H3A 2T8}

\date{\today}

\begin{abstract}
Experimental data for central collisions of $^{129}$Xe on $^{119}$Sn at beam energies of (a) 32 MeV/nucleon, (b) 39 MeV/nucleon, (c) 45 MeV/nucleon and
(d) 50 MeV/nucleon are compared with results calculated using a hybrid model.  We use a transport model (BUU) to obtain the excitation energy per nucleon in the center of mass of the multifragmenting system. The canonical thermodynamic model is then used to determine the temperature which would lead to this excitation energy.  With this temperature we use the canonical thermodynamic model to calculate various experimental data such as multiplicities of different composites, probability distribution of the largest cluster etc.  Agreement with data establishes the validity of the model.
\end{abstract}


\maketitle
\section{Introduction}
Nuclear multifragmentation is an important phenomenon, the study of which can reveal reaction mechanism in heavy ion collisions at intermediate and high energies \cite{Moretto,Gross1,Jacob,Cole,Mallik102}. Central collision fragmentation reactions around fermi energy domain are extensively used for producing neutron rich isotopes and for studying nuclear liquid gas phase transition.\\
Different theoretical models have been already developed for throwing light on the nuclear multifragmentation reaction and for explaining relevant experimental data. The models are mainly classified into two categories: (i) Dynamical models \cite{Dasgupta,Ono,Hartnack} and (ii) Statistical models \cite{Das1,Bondorf1,Gross2,Randrup,Brohm,Raduta,Lacroix}. The dynamical models are based on more microscopic calculations where the time evolution of projectile and target nucleons are studied. In statistical models the clusterization technique is nicely incorporated but the disadvantage of statistical model is that the calculation are started by assuming some initial conditions (like temperature, excitation energy, freeze-out volume, fragmenting source size etc.). These condition are either parameterized or obtained from some experimental observables.\\
In this work we develop a hybrid model for explaining multifragmentation reaction around fermi energy domain. We treat central collision only.  Initially the excitation of the colliding system is calculated by using dynamical Boltzmann-Uehling-Uhlenbeck (BUU) approach \cite{Dasgupta,Mallik9} with proper consideration of pre-equilibrium
emission. Then the fragmentation of this excited system is calculated by Canonical Thermodynamical model (CTM) \cite{Das1}. The decay of excited fragments, which are produced in multifragmentation stage is calculated by an evaporation model \cite{Mallik1} based on Weisskopf theory. Different observables like charge distribution, largest cluster distribution etc. are calculated by using this hybrid model for $^{129}$Xe+$^{119}$Sn reaction at different projectile energies and compared with experimental data \cite{Hudan} The idea of setting the initial conditions for a statistical model from a dynamical model is of course not new; see for example Barz et al \cite{Barz}. In many Statistical Model of Multifragmentation (SMM), the initial conditions are fixed by some measured data. In our hybrid model the initial conditions for the thermodynamical model are set up almost entirely by the transport model calculation.\\
The concept of temperature is quite familiar in heavy ion collision and it is a better observable (compared to energy) for studying liquid gas phase transition. One standard way of extracting temperatures is the Albergo formula \cite{Albergo}, where temperature is calculated from the measured isotopic yields (i.e. cold fragments). Another common technique for obtaining temperature is to measure the kinetic energy spectra of emitted particles. But in this method, the effect of sequential decay from higher energy states, Fermi motion, pre-equilibrium emission etc complicate the scenario of temperature measurement. Our hybrid model calculation is an alternative method for deducing the freeze-out temperature and it bypasses all such problems.\\
\begin{figure*}
\includegraphics[width=16cm,keepaspectratio=true,clip]{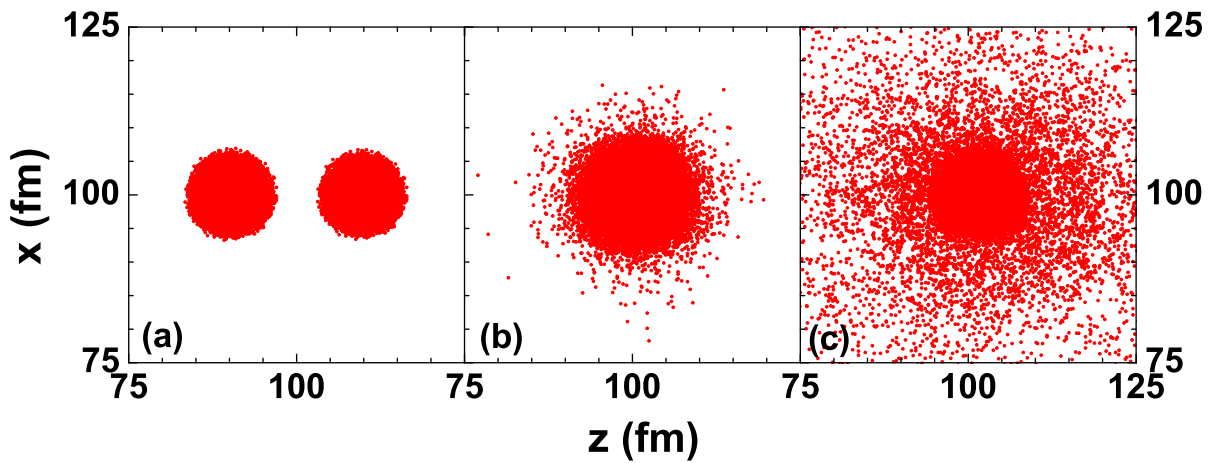}
\caption{(Color online) Evolution of test particles at (a) 0 fm/c, (b) 75 fm/c and (c) 200 fm/c in center of mass frame for 45 MeV/nucleon $^{129}$Xe on $^{119}$Sn reaction.}
\label{fig1}
\end{figure*}
\section{Basics of the dynamical model}
The hybrid model consist of three different stages: (i) Initial condition determination by dynamical BUU calculation, (ii) fragmentation by canonical thermodynamical model and (iii) decay of excited fragments by evaporation model.\\\
We start our calculation when two nuclei in their respective ground states approach each other with specified velocities. The mean field potential energy
density is taken from Lenk-Pandharipande \cite{Lenk}:
\begin{equation}
v(\rho(\vec{r}))=\frac{A}{2}\rho^2(\vec{r})+
\frac{B}{\sigma+1}\rho^{\sigma+1}(\vec{r}) +\frac{c\rho_0^{1/3}}{2}
\frac{\rho(\vec{r})}{\rho_0}\nabla_r^2[\frac{\rho(\vec{r})}{\rho_0}]
\end{equation}
where first two term represents zero range Skyrme interaction and the derivative term does not affect nuclear matter properties but in a finite system it produces quite realistic diffuse surfaces and liquid drop binding energies. This can be achived for $A=$-2230.0 MeV $fm^3, B$=2577.85 MeV $fm^{7/2}, \sigma=$7/6, $\rho_0=0.16$ and $c$=-6.5 MeV$fm^{5/2}$ \cite{Lenk}. We first construct Thomas-Fermi solutions for ground states \cite{Lee}. The Thomas-Fermi phase space distribution will then be modeled by choosing test particles with appropriate positions and momenta using Monte Carlo. Each nucleon is represented by 100 test particles ($N_{test}=100$). We begin Boltzmann-Uehling-Uhlenbeck (BUU)  model calculation to get the excitation of the fragmenting system.
In the center of mass frame the test particles of the projectile and the target nuclei (in their Thomas-Fermi ground state) are boosted towards each other.  The test particles move in a mean-field $U(\rho(\vec{r}))$ (generated by the potential energy density of Eq.(1)) and will occasionally suffer two-body collisions when two of them pass close to each other and the collision is not blocked by Pauli principle. The mean-field propagation is done using the lattice Hamiltonian method which conserves energy and momentum very accurately \cite{Lenk}. Two body collisions are calculated as in Appendix B of ref. \cite{Dasgupta}, except that pion channels are closed, as there will not be any pion production in this energy region. Positions and momenta of the test particles are updated after each time steps ($\Delta t$) by the equations
\begin{eqnarray}
\frac{d\vec{p}_i}{dt}&=&-\nabla_r U(\rho(\vec{r}_i),t)\nonumber\\
\frac{d\vec{r}_i}{dt}&=&\vec{v}_i\nonumber\\
&& i=1,2,.....,(A_p+A_t)N_{test}
\end{eqnarray}

\section{Excitation Energy Determination}

We can calculate the excitation energy ($E^*$) from
projectile beam energy ($E_{beam}$) by direct kinematics by assuming that the projectile and the
target fuse together.  In that case the excitation energy is
$E^*=A_pE_{beam}/(A_p+A_t)$ where $A_p$ and $A_t$ are projectile and target
masses respectively.  This value is too high as a measure of the excitation
energy of the system which multifragments.  Pre-equilibrium particles
which are not part of the multifragmenting system carry off a significant
part of the energy.

\begin{figure}[b]
\includegraphics[width=3.0in,height=3.0in,clip]{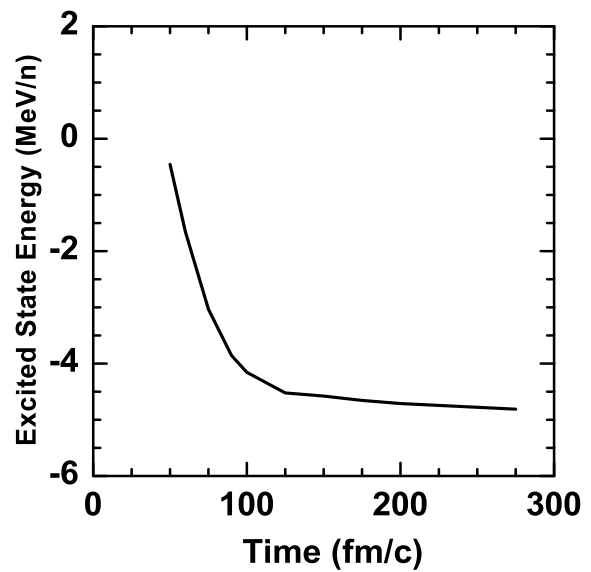}
\caption{Variation of energy of the central dense region (containing $80\%$ of total test particles) with time obtained from dynamical BUU calculation for $^{129}$Xe on $^{119}$Sn reaction at 45 MeV/nucleon.}
\label{fig2}
\end{figure}
\begin{figure}[t]
\includegraphics[width=\columnwidth,height=2.5in,clip]{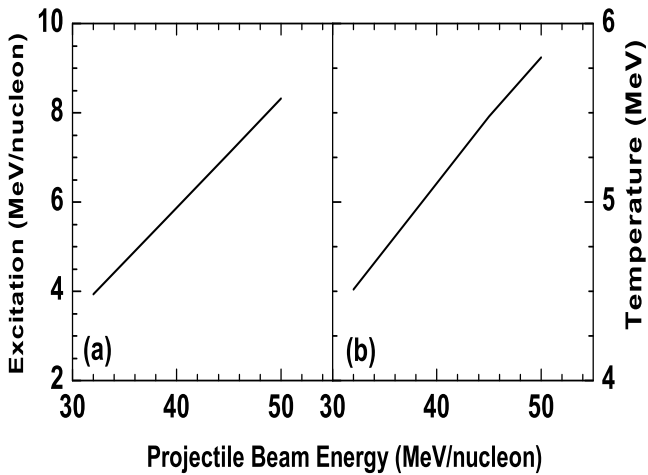}
\caption{Left Panel indicates the variation of excitation energy per nucleon with projectile beam energy per nucleon obtained from dynamical BUU model. The Canonical Thermodynamical Model (CTM) can calculate average excitation energy per nucleon for a given freeze-out temperature, mass number and charge. Therefore to know the required freeze-out temperature corresponding to each excitation (obtained from BUU calculation) CTM is used. The variation of freeze-out temperature with projectile beam energy is shown in the right panel.}
\label{fig3}
\end{figure}
To get a better measure of excitation of the fragmenting system we need to do a BUU calculation where the pre-equilibrium particles can be identified and can be taken out to calculate excitation energy per nucleon. We exemplify our method with central collision reactions $^{129}$Xe+$^{119}$Sn at projectile beam energy $45$ MeV/nucleon. Initially the center of $^{129}$Xe and $^{119}$Sn are kept at (100fm, 100fm, 90fm) and (100fm, 100fm, 110fm) respectively and and they are boosted towards each other along $z$ direction. Fig. 1 shows the test particles at t=0 fm/c (when the nuclei are separate), 75 fm/c (the time when violent collisions occur) and 200 fm/c (almost all collisions are completed). From the figure it is clear that for t=200 fm/c some test particles are far distant from the central dense region. These fit the category of pre-equilibrium emission. In different multifragmentation experiments, it is observed that after pre-equilibrium emission, $75\%$ to $80\%$ of the total mass creates the fragmenting system \cite{Xu,Frankland,Verde}. Hence we choose the test particles which create $80\%$ of the total mass (i.e. $A_0=198$) from the most central dense region. Knowing the momenta of selected test particles the kinetic energy is calculated and from the positions of these selected test particles the potential energy is calculated by using Eq. 1. By adding kinetic and potential energy the energy of the fragmenting system is obtained.  Fig. 2 shows the variation of excited state energy of the central dense region (i.e. $80\%$ of the total  test particles) with time. Here total energy is always constant but as time progresses, pre-equilibrium particles having high kinetic energy, are escaping from the central dense region, therefore the energy of the central dense region is decreasing. It is clear that after t=100 fm/c, the energy becomes independent of time. Hence, we can stop BUU calculation at any time after t=100 fm/c and consider the corresponding energy as excited state energy. To get the excitation we need to know the ground state of the fragmenting system. For this we use the Thomas Fermi method for a spherical nucleus of mass $A=198$ ($80\%$ of $^{129}$Xe+$^{119}$Sn mass). Subtracting ground state energy from the calculated energy above the excitation energy is obtained.

\begin{figure}[t]
\includegraphics[width=\columnwidth,keepaspectratio=true,clip]{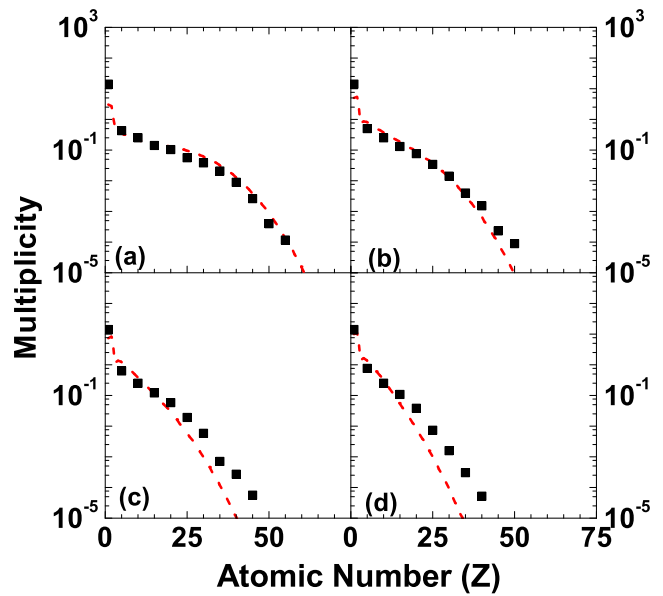}
\caption{(Color online) Theoretical charge distribution (red dotted lines) for $^{129}$Xe
on $^{119}$Sn reaction at (a) 32 MeV/nucleon (b) 39 MeV/nucleon
(c) 45 MeV/nucleon and (d) 50 MeV/nucleon. The experimental data are shown by
black squares.}
\label{fig4}
\end{figure}
\section{Computations with the statistical model: Extraction of Temperature}
We have described above how from BUU we extract the mass, charge and the excitation energy of the fragmenting system. Our next task is to obtain the freeze-out temperature. The canonical thermodynamic model (CTM) \cite{Das1} can be used to calculate average excitation per nucleon for a given temperature, charge and mass.  Getting an excitation energy for a given temperature, mass and charge is described in detail in \cite{Das1}. We do the exploration for each beam energy. We will not repeat the formulae of CTM here but just mention that apart from neutrons and protons the following composites are included in CTM breakup.  We include deuteron,triton, $^3He$, $^4He$ and for heavier nuclei we include a ridge along the line of stability. The composites that follow from CTM will further decay by evaporation.  The details of how we do it can be found in \cite{Mallik1}.
\section{Results}
We have done calculations for the same $^{129}$Xe+$^{119}$Sn pair for projectile beam energies 32, 39, 45 and 50 MeV/nucleon. In each case, we have stopped the time evolution at $t=200$ fm/c, and selected $80\%$ of the total mass from central dense region for calculating the excited state energy. Then subtracting the ground state energy the excitation is obtained. The variation of calculated excitation energy with projectile beam energy is shown in the left diagram of Fig. 3. From this excitation energy we find out the corresponding freeze-out temperature.  Thus the freeze-out temperature for a given beam energy is obtained. This is plotted on the right side of Fig.3.\\
\begin{figure}[h!]
\includegraphics[width=\columnwidth,keepaspectratio=true,clip]{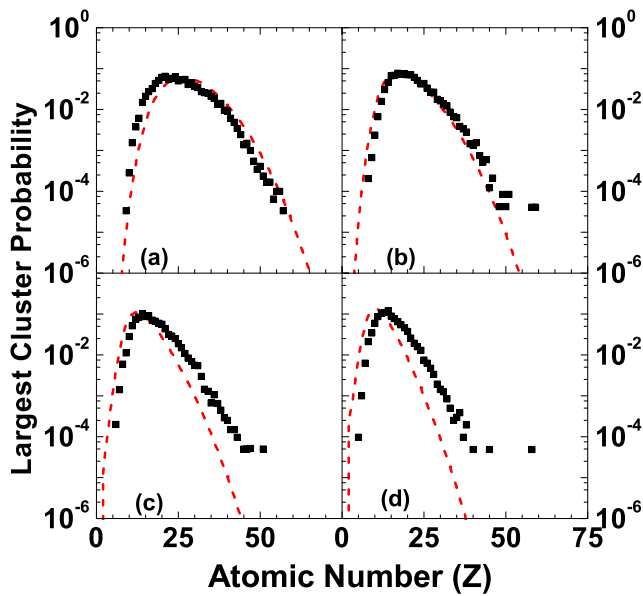}
\caption{(Color online) Theoretical largest cluster probability distribution (red dotted lines) for $^{129}$Xe on $^{119}$Sn reaction at (a) 32 MeV/nucleon (b) 39 MeV/nucleon (c) 45 MeV/nucleon and (d) 50 MeV/nucleon. The experimental data are shown by black squares.}
\label{fig5}
\end{figure}
To check the accuracy of our model, we have compared the theoretical results with experimental data. Fig. 4 shows the comparison of charge distribution at projectile beam energies 32, 39, 45 and 50 MeV/nucleon. With the increase of energy (i.e. increase of temperature), fragmentation is more, therefore multiplicities of higher fragments gradually decrease. Fig. 5 represents the largest cluster probability distribution at different energies. Since with the increase of energy breaking increases, the peak of the largest cluster probability distribution shifts towards the lower atomic number side and the width of the distribution gradually decreases. The variation of average charge of largest cluster $\langle Z_{Largest}\rangle$ with projectile beam energy is shown in Fig. 6. In each case nice agreement between theoretical result and experimental data is obtained.\\
\begin{figure}[t]
\includegraphics[width=2.5in,height=2.5in,clip]{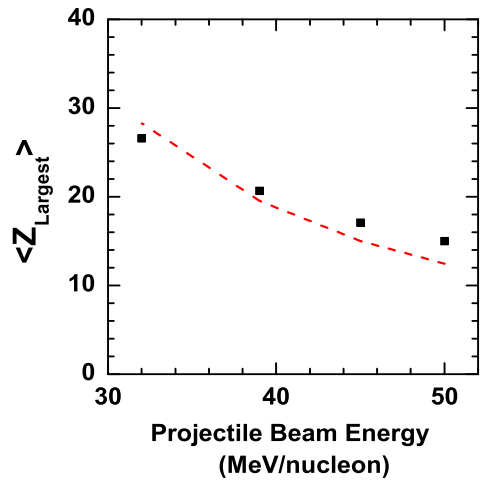}
\caption{(Color online) Variation of average size of largest cluster with projectile beam energy obtained from hybrid model calculation (red dotted lines) for $^{129}$Xe on $^{119}$Sn reaction. The experimental data are shown by black squares.}
\label{fig6}
\end{figure}
\section{Discussion}
In this work we do a BUU calculation to get the excitation energy of the multifragmenting system produced in the central collision reactions around Fermi energy domain, then do a CTM exploration to locate the temperature which will give this excitation.  Once this temperature is fixed CTM is used to fit available experimental data. The agreement with data is pleasing. This work complements the work we did where we fitted the data obtained from
the decay of projectile like fragments at energies in the limiting fragmentation region.  There \cite{PLF} we fitted the data using CTM with an assumed temperature profile first and later \cite{Mallik9} we showed that the temperature profile is obtainable from BUU calculations. Our present calculations are not prohibitively computer intensive.  One virtue of these calculations is that equation 1 leads to reasonable values of binding energy of finite nuclei (even in Thomas-Fermi approximation) and realistic diffuse surface without having to supplement the zero range Skyrme interaction with a finite range interaction.
Vlasov propagation for large nuclei when finite range interaction is present is very computer intensive.  The other pleasing aspect is that the lattice Hamiltonian method \cite{Lenk} gives remarkable accuracy in total energy and total momentum conservation in these calculations.\\
For fragmenting system, we adopted the value of $80\%$ of total mass from the experimental papers quoted in our paper. But our results (see Fig. 6) show that this was a reasonable choice. We show a plot of $\langle Z_{Largest}\rangle$ which agrees fairly well with data. Now $\langle Z_{Largest}\rangle$ depends upon the size of the fragmenting system as well as the temperature of the fragmenting system. The larger the fragmenting system, the larger is the $\langle Z_{Largest}\rangle$. The higher the temperature, the smaller is $\langle Z_{Largest}\rangle$. Now the temperature also depends upon what percentage of nucleons are left out as pre-equilibrium particles. The value $80\%$ we choose gives a combination of temperature and fragmenting mass that seems to be just about right. One could do a detailed best "fit" but this was not attempted.\\
What we presented in this work did not involve any radial flow. One reason is that the collision energy being only about 50 MeV/nucleon, the initial compression is small so any radial flow must also be small. The best signature for radial flow will be in the velocity distribution but we are only calculating multiplicity distribution. Neither CTM nor SMM can incorporate radial flow easily but in Lattice gas model, where flow is easily incorporated. it was found that even for significant radial flow, multiplicity distributions are hardly affected \cite{Das_LCG}.
\section{Acknowledgements}
This work was supported in part by Natural Sciences and Engineering Research Council of Canada.


\begin{thebibliography}{999}
\bibitem{Moretto} L. G. Moretto and G. J. Wozniak, Annu. Rev. Nucl. Part. Sci. {\bf 43}, 379 (1993).
\bibitem{Gross1} D. H. Gross, Eur. Phys. J. A. {\bf 30}, 293 (2006).
\bibitem{Jacob} B. V. Jacak , Nucl. Phys. A 488, 325c (1988)
\bibitem{Cole} A. J. Cole et. al. , Eur. J. Phys. {\bf 18}, 425 (1997).
\bibitem{Mallik102} G. Chaudhuri, S. Mallik and S. Das Gupta,  Pramana J. Phys. {\bf82}, 907 (2014).
\bibitem{Dasgupta} G. F. Bertsch and S. Das Gupta, Phys. Rep {\bf 160}, 189 (1988).
\bibitem{Ono} A. Ono and H. Horiuchi, Prog. Part. Nucl. Phys {\bf 53}, 501 (2004).
\bibitem{Hartnack} C. Hartnack et al., Eur. phys. J. A{\bf 1} 151 (1998)
\bibitem{Das1} C. B. Das, S. Das Gupta et al., Phys. Rep {\bf 406}, 1 (2005).
\bibitem{Bondorf1} J. P. Bondorf et al., Phys. Rep. {\bf 257}, 133 (1995).
\bibitem{Gross2} D. H. Gross, Phys. Rep {\bf 279}, 119 (1997).
\bibitem{Randrup} J. Randrup and S. E. Koonin , Nucl. Phys. A356, 223(1981)
\bibitem{Brohm} T. Brohm and K-H Schmidt, Nucl. Phys. A569, 821(1994)
\bibitem{Raduta} Al. H. Raduta and  Ad. R. Raduta, Phys. Rev. {\bf C 55}, 1344 (1997).
\bibitem{Lacroix} D. Lacroix, A. Van Lauwe, and D. Durand, Phys. Rev. {\bf C 69}, 054604 (2004).
\bibitem{Mallik9} S. Mallik, S. Das Gupta and G. Chaudhuri, Phys. Rev. {\bf C 89}, 044614 (2014).
\bibitem{Mallik1} G. Chaudhuri and S. Mallik, Nucl. Phys. {\bf A 849}, 190 (2011).
\bibitem{Barz} H. W. Barz et al., Nucl. Phys. A561, 466(1993)
\bibitem{Albergo} S. Albergo et al., Il Nuovo Cimento {\bf 89}, A1 (1985).
\bibitem{Hudan} S. Hudan, A. Chbihi et. al. , Phys. Rev. {\bf C 67}, 064613 (2003).
\bibitem{Lenk} R. J. Lenk and V. R. Pandharipande, Phys. Rev. {\bf C 39}, 2242 (1989).
\bibitem{Lee} S. J. Lee, H. H. Gan, E. D. Cooper and S. Das Gupta, Phys. Rev. {\bf C 40}, 2585 (1989).
\bibitem{Xu} H. S. Xu et al., Phys. Rev. Lett. {\bf 85}, 716 (2000).
\bibitem{Frankland} J. D. Frankland et al., Nucl. Phys. A 649, 940 (2001)
\bibitem{Verde} G. Verde , Braz. Jour. of Phys. 37, 885 (2007)
\bibitem{Myers} W. D. Myers, Nucl. Phys. A296,177(1978)
\bibitem{Weisskopf} V. Weisskopf, Phys. Rev. {\bf 52}, 295 (1937).
\bibitem{PLF} S. Mallik, S. Das Gupta and G. Chaudhuri, Phys. Rev. {\bf C83}, 044612 (2011)
\bibitem{Das_LCG} C. B. Das, L. Shi and S. Das Gupta, Phys. Rev. {\bf C70}, 064610 (2004)
\end{thebibliography}
\end{document}